# Mutual synchronization of nano-oscillators driven by pure spin current

S. Urazhdin[1], V. E. Demidov[2*], R. Cao[1], B. Divinskiy[2], V. Tyberkevych[3], A. Slavin[3], A. B. Rinkevich[4], and S. O. Demokritov[1,4]

[1]Department of Physics, Emory University, Atlanta, GA 30322, USA

[2]Institute for Applied Physics and Center for Nanotechnology, University of Muenster, 48149 Muenster, Germany

[3]Department of Physics, Oakland University, Rochester, MI 48309, USA

[4]M.N. Miheev Institute of Metal Physics of Ural Branch of Russian Academy of Sciences, Yekaterinburg 620041, Russia

We report the experimental observation of mutual synchronization of magnetic nano-oscillators driven by pure spin current generated by nonlocal spin injection. We show that the oscillators efficiently synchronize due to the direct spatial overlap of the dynamical modes excited by spin current, which is facilitated by the large size of the auto-oscillation area inherent to these devices. The synchronization occurs within an interval of the driving current determined by the competition between the dynamic nonlinearity that facilitates synchronization, and the short-wavelength magnetic fluctuations enhanced by the spin current that suppress synchronization. The demonstrated synchronization effects can be utilized to control the spatial and spectral characteristics of the dynamical states induced by spin currents.

The recently demonstrated magnetic nano-oscillators driven by pure spin currents[1,2] have attracted a significant attention due to their promise for applications in spintronics and magnonics[3-14]. By utilizing pure spin currents that do not require electrical current flow through the active magnetic layer, one can reduce undesirable Joule heating in spintronic nano-circuits, minimize the effects of spatially non-uniform Oersted fields of the driving current that can complicate the current-induced magnetic dynamics, and develop spin-torque nano-devices based on low-loss insulating magnetic materials.

The majority of the previously demonstrated spin-current oscillators utilized spin currents generated by the spin-Hall effect (SHE)[15,16]. However, devices based on another mechanism for pure spin current generation – nonlocal spin injection (NLSI)[17,18] – provide several important advantages. In addition to the possibility to generate highly coherent confined magnetization oscillation[10,12], these devices enable generation of propagating spin waves[14], which is facilitated by their simple and flexible geometry, as well as by the large size of the auto-oscillation area inherent to the NLSI mechanism. Since the NLSI mechanism enables coherent room-temperature oscillations in extended magnetic films with a large size of the auto-oscillation area, it provides an opportunity to implement efficient mutual coupling of different oscillators by direct overlap of their oscillation regions. This opens the possibility for the implementation of ensembles of mutually synchronized nano-oscillators that can exhibit improved oscillation coherence and increased generated microwave power or spin wave intensity.

Mutual synchronization has been extensively studied in ensembles of traditional nano-oscillators operating with spin-polarized electric currents, where synchronization of up to five devices has been demonstrated[19-23]. The geometric flexibility of the oscillators driven by pure spin current is expected to facilitate the implementation of synchronized ensembles. However, only synchronization to external microwave signals has been experimentally demonstrated for these devices[4], while their mutual synchronization has been studied only theoretically[24,25].

In this Letter, we report the experimental observation of mutual synchronization of two spin-current nano-oscillators. We show that efficient synchronization can be easily achieved in devices based on the NLSI spin-current generation mechanism. We take advantage of the flexible layout of these devices enabling optical access to the active area to directly image the current-induced auto-oscillations by magneto-optical spectroscopy. We show that the auto-oscillation regions of two devices separated by a distance of 400 nm strongly overlap, resulting in their efficient interaction. We also identify the mechanisms responsible for onset of synchronization and its loss at large driving currents, and show that the interval of synchronization currents can be increased by increasing the static magnetic field.

The schematic of the experiment is shown in Fig. 1. The studied structure consists of two NLSI nano-oscillators placed at a distance of 400 nm from each other. The oscillators are formed by two 60 nm circular nanocontacts on an extended multilayer that consists of a 5 nm thick Permalloy (Py) active magnetic layer separated from the 8 nm thick CoFe spin injector by a 20 nm thick layer of Cu. The driving electric current $I$ is injected into the multilayer through both nanocontacts simultaneously, and is drained to the side electrodes located at the distance of 2 μm from the nanocontact pair (not shown in Fig. 1). Because of the large difference in the resistivities of the materials comprising the device, most of the current is drained through the Cu layer, while only 3% of the current is shunted through the active Py layer[14]. The red arrow in Fig. 1 shows the corresponding flow of electrons. The injected electrons become spin polarized due to the spin-dependent scattering in CoFe and at the Cu/CoFe interface[26], resulting in the spin accumulation in Cu above the nanocontacts. Spin diffusion away from this region produces a spin current flowing into the Py layer, exerting spin transfer torque (STT)[27,28] on its magnetization. The magnetizations of both CoFe and Py layers are aligned with the saturating static in-plane magnetic field $H_0$. For the positive driving electric currents, as defined in Fig. 1, the magnetic moment carried by the spin current is antiparallel to the magnetization of the Py layer. The resulting STT reduces the effective dynamic magnetic damping of Py. When damping is completely compensated by the spin current, the magnetization of the Py layer starts to auto-oscillate in the region of the strongest spin-current injection above the nanocontacts[10].

We detect the spin current-induced oscillations by micro-focus Brillouin light scattering (BLS) spectroscopy[29]. The probing light with the wavelength of 532 nm generated by a single-frequency laser is focused on the surface of the Py film into a diffraction-limited spot. The probing light interacts with the magnetic oscillations in Py, resulting in its modulation. The spectral satellites caused by the modulation are analyzed by a tandem Fabry-Perot interferometer. The intensity of the resulting BLS signal is proportional to the intensity of magnetization oscillations at the position of the probing spot.

Figure 2 shows the spatial maps of the spin current-induced magnetization dynamics, recorded by rastering the probing spot in the two lateral dimensions while measuring the BLS intensity. The BLS map acquired at $I$=12.5 mA (Fig. 2(a)) clearly shows auto-oscillations in the Py film above the nanocontact labeled "A", while the nano-oscillator labeled "B" remains in the sub-critical regime. This asymmetry is likely caused by a small difference in the diameters of the two nanocontacts, resulting in a slight difference between their auto-oscillation onsets. This imbalance fortuitously allows us to estimate the size of the auto-oscillation area. We first fitted the *x*- and *y*-sections of the BLS map by the Gaussian function. These profiles represent a convolution of the actual dynamic magnetization intensity distribution with the profile of the laser spot characterized by the diffraction-limited diameter of about 250 nm[1,29]. By performing a deconvolution, we determined that the full width of the auto-oscillation area at half-maximum intensity is 350 nm in the direction perpendicular to the static field, and 250 nm in the direction parallel to it. These values are in a good agreement with the results for the standalone NLSI oscillators[10]. Note that the size of the auto-oscillation region in the direction along the line connecting the two nanocontacts is close to the separation between them, which can be expected to result in a significant spatial overlap of the oscillating magnetization regions when both devices are in the auto-oscillation regime. Indeed, the data obtained at *I*=20 mA (Fig. 2(b)) show two spatially overlapping oscillating magnetization regions centered on the respective point contacts.

To analyze the interaction of the two nano-oscillators caused by the overlap of their auto-oscillation regions, we independently measured their oscillation characteristics. This was accomplished by placing the probing laser spot at the location of the corresponding nanocontact, and recording the BLS spectrum at this position. The spectra measured for different driving dc currents are presented in Fig. 3. The device B starts to oscillate at a frequency which is higher than the frequency of device A by Δ*f*=0.4 GHz (*I*=16.5 mA in Fig. 3). This difference, which is likely caused by the slightly different diameters of the nanocontacts, provides an opportunity to identify different regimes associated with the interaction between the oscillators. The value of $\Delta f$ slightly increases with increasing current, as shown for $I$=17.5 mA in Fig.3. The oscillation peaks abruptly jump at a current $I_{SYN}$=18 mA and their frequencies become equal, indicating mutual synchronization of the two oscillators[30] ($I$=18.5 mA in Fig. 3). As the current is further increased, both peaks shift to smaller frequencies due to the nonlinear frequency shift, but their frequencies remain equal ($I$=23 mA in Fig. 3). The auto-oscillation peaks start to gradually diverge above a current $I_L$=23 mA, indicating the loss of synchronization, as illustrated for $I$=27 mA and 30 mA in Fig. 3.

By fitting the recorded spectra with the Gaussian function, we quantified the dependencies of the frequencies and the amplitudes of the two oscillators on dc current (Fig.4). The most prominent feature in the data is a significant abrupt jump of the oscillation frequencies at the synchronization onset current $I_{SYN}$, as marked by the dashed line in Fig. 4(a). The jump is significantly larger for device B, which starts to oscillate at larger currents, than for device A. The synchronization onset also coincides with the maximum amplitude of oscillations reached by device B (Fig. 4(b)).

Note that the oscillation characteristics of device B are not significantly different from those of device A. On the contrary, aside from the effects of synchronization and the difference in the onset current, both oscillators exhibit very similar dependencies of frequency and amplitude on current. Specifically, the frequencies of both oscillators remain approximately constant at currents within 4-5 mA from the respective onset currents (Fig. 4 (a)), while their amplitudes rapidly increase over the same current ranges (Fig. 4(b)). The amplitudes peak at approximately the same level, and then start to gradually decrease. The behaviours at large currents can be associated with the onset of nonlinear coupling between the auto-oscillating mode and the incoherent magnetization fluctuations, which are known to be strongly enhanced by the spin current[31]. In particular, this enhancement results in the onset of the nonlinear damping resulting in the reduction of the amplitude of auto-oscillations.

To understand the mechanisms controlling the synchronization, we note that the dynamical nonlinearity is expected to play a significant role in this process[32]. The nonlinearity can be characterized by the slope of the dependence of the auto-oscillation frequency on current. At currents slightly above the onset, where the auto-oscillation amplitude is still small, the frequency of the oscillators is weakly dependent on current, indicating a weakly nonlinear oscillation regime (Fig.4(a)). The frequency starts to redshift at larger currents, indicating a transition to a strongly nonlinear regime. The synchronization occurs when such a transition occurs for device B, when device A is already in the nonlinear regime. Thus, the synchronization occurs when both oscillators enter the strongly nonlinear regime, in agreement with the theoretical models of synchronization in STNO[32]. We emphasize that, in the strongly nonlinear regime, the frequency redshift is determined by both the intensity of the long-wavelength auto-oscillations, which can be directly detected in the experiment, and the intensity of short-wavelength fluctuations, which are strongly coupled to the auto-oscillation mode but are not directly detectable by the BLS. Therefore, the nonlinear redshift remains strong even at large currents, where the intensity of the auto-oscillations noticeably decreases.

In contrast to the synchronization onset, the loss of synchronization at large current is not abrupt, but is characterized by a gradual increase of the difference $\Delta f$ between the frequencies of the two oscillators. This indicates that the synchronization is not completely lost at large currents, but is instead increasingly disturbed with increasing *I*. We argue that the only mechanism capable of gradually disturbing the synchronization is the gradual enhancement of magnetization fluctuations by the spin current. Because of the nonlinear coupling among different dynamical modes, enhanced fluctuations affect the phase stability of the oscillators, causing a partial loss of their synchronization. The enhancement of the fluctuations, which do not directly contribute to the BLS signal, is apparent from the large-current oscillation characteristics shown in Fig.4: the frequency redshift observed at large current, when the amplitude of the auto-oscillations saturates and then decreases, can be explained only by the decrease of the effective magnetization due to the increase of the total amplitude of fluctuations. The dominant contribution of the fluctuations to the loss of synchronization is also in agreement with the results of Ref. 4, which showed that the frequency interval of synchronization of SHE-based spin-current oscillators to external microwave signals is limited by the magnetization fluctuations enhanced by the spin current.

The dependence of the oscillation characteristics on the applied field (Fig.5) provides further insight into the mechanisms of synchronization. Figure 5(a) shows the current dependence of the difference $\Delta f$ between the auto-oscillation frequencies of the two oscillators, at $H_0$=1000 Oe and 1400 Oe. Both dependencies exhibit a jump to $\Delta f$=0 associated with the onset of synchronization, starting from the current $I_{SYN}$=18 mA at $H_0$=1000 Oe, and $I_{SYN}$=16 mA at $H_0$=1400 Oe. The frequencies of the oscillators remain equal within a certain interval of *I* (hatched area in Fig. 5(a) for $H_0$=1000 Oe), followed by the loss of synchronization marked by a gradual increase of $\Delta f$ starting from a well-defined current $I_L$=23 mA at $H_0$=1000 Oe, and $I_L$=26 mA at 1400 Oe.

The field dependences of $I_{SYN}$, $I_L$, and of the auto-oscillation onset current $I_{OSC}$ of oscillator B are summarized in Fig. 5(b). The synchronization current $I_{SYN}$ monotonically decreases with increasing field, while the synchronization loss current increases, resulting in an overall significant increase of the synchronization current interval. In contrast, the auto-oscillation onset current exhibits a weak non-monotonic dependence on field. Thus, we can conclude that the variations of $I_{SYN}$ and $I_L$ are not associated with a simple rescaling of the characteristic oscillation currents.

The observed dependencies clearly indicate that the synchronization is facilitated by the field increase. To explain this behaviour, we note that the dynamic nonlinearity favours synchronization, while the incoherent fluctuations suppress synchronization. The fluctuations are not expected to significantly depend on the field, because they are dominated by the large phase volume of short-wavelength modes whose spectral characteristics are mostly determined by the exchange interaction, not by the modest static field. In contrast, the nonlinearity of the dynamical states of in-plane magnetized films significantly increases with increasing static field[33]. Since the nonlinearity facilitates synchronization, this results in an increase of the synchronization current interval with increasing field.

In conclusion, we have demonstrated that the spin current-driven nano-oscillators are capable of efficient mutual synchronization over a wide range of experimental parameters. We showed that the observed synchronization behaviors can be qualitatively understood in terms of a competition between the dynamical nonlinearity and magnetic fluctuations enhanced by the spin current. Additional insights into these behaviors, as well as the clarification of the contributions to the synchronization of other mechanisms such as exchange of propagating spin waves, coupling mediated by spin currents in Cu, and the dipolar interaction, can be gained from the rigorous theoretical analysis incorporating nonlinear interactions among the different strongly driven spin-wave modes of the magnetic system, which will likely provide directions for the further improvement of the dynamical characteristics of interacting spin-current oscillators. The NLSI-based devices are particularly amenable to such optimization thanks to their flexible geometry, making them promising for the spintronic and magnonic applications.

This work was supported in part by the Deutsche Forschungsgemeinschaft, NSF Grants ECCS-1305586, ECCS-1509794, and the program Megagrant № 14.Z50.31.0025 of the Russian Ministry of Education and Science.

**FIGURE CAPTIONS**

Fig. 1 (color online) Schematic of the experiment.

Fig.2 (color online) Color-coded two-dimensional maps of the magnetization oscillations induced by pure spin current in the Py film, and their cross-sections along the line connecting the nanocontacts. The maps were recorded at field $H_0$=1000 Oe and currents $I$=12.5 (a) and 20 mA (b). The white circles in the maps and the dashed lines in the cross-sections mark the positions of the nanocontacts.

Fig. 3 (color online) Normalized BLS spectra of auto-oscillations recorded at different dc currents, as labeled, by placing the probing laser spot above the nanocontact A ($x$=0.2 µm) and B ($x$=-0.2 µm). Symbols: experimental data, curves: Gaussian fits. The data were obtained at $H_0$=1000 Oe. The spectral width of the peaks is determined by the limited frequency resolution of the BLS technique.

Fig. 4 (color online) Current dependencies of the auto-oscillation frequencies (a) and the amplitudes (b) for devices A (solid symbols) and B (open symbols). The vertical dashed line marks the onset of synchronization. The data were obtained at $H_0$=1000 Oe.

Fig. 5 (color online) (a) Current dependence of the difference between the auto-oscillation frequencies of the oscillators A and B for two magnitudes of the static magnetic field, as labeled. The hatched area marks the synchronization interval at $H_0$=1000 Oe. (b) Field dependencies of the auto-oscillation onset current $I_{\mathrm{OSC}}$, the synchronization onset current $I_{\mathrm{SYN}}$, and the current $I_{\mathrm{L}}$ at which the synchronization is lost.

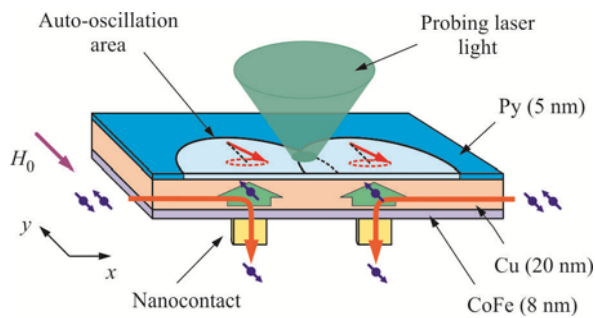

Fig. 1

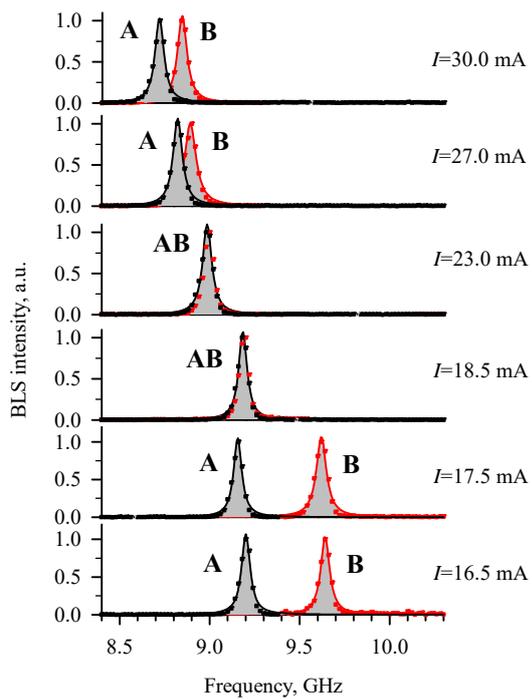

Fig. 3

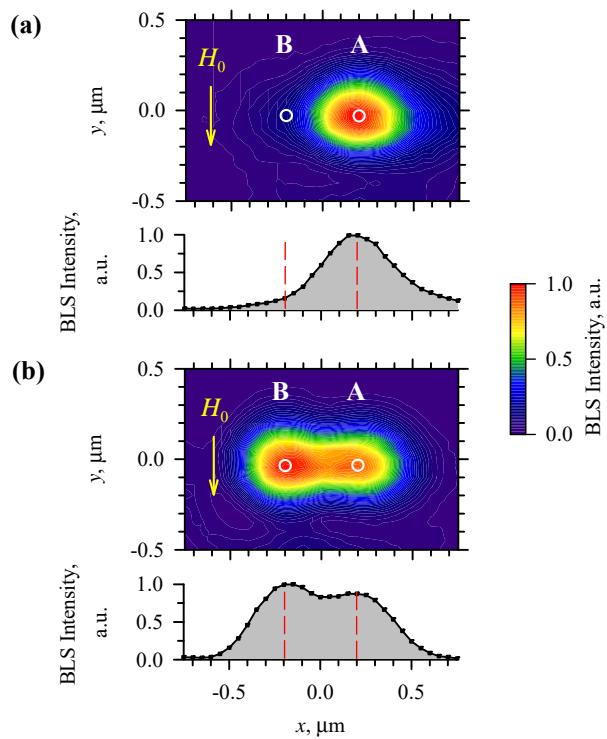

Fig. 2

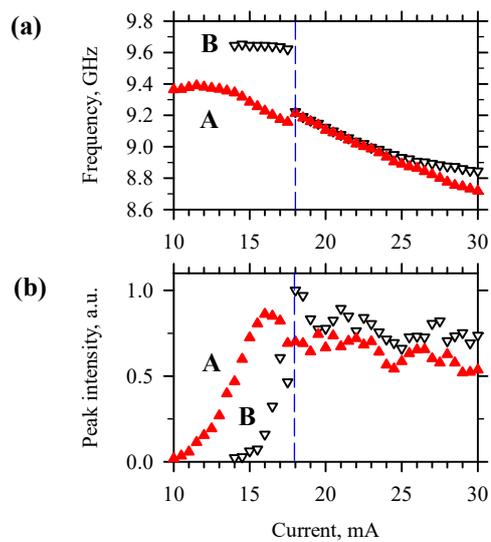

Fig. 4

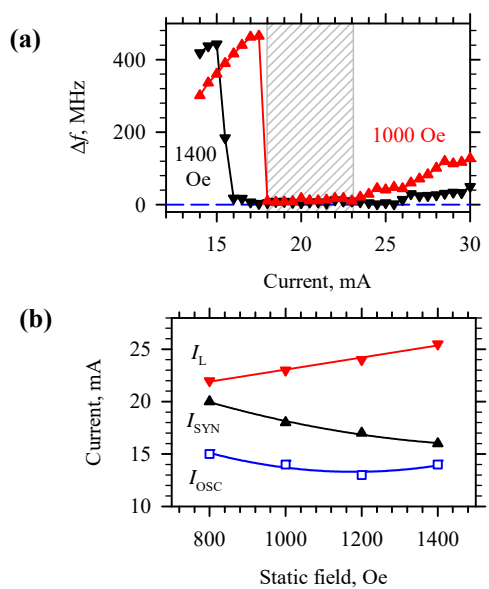

Fig. 5